\documentclass{article}

\usepackage{PRIMEarxiv}

\usepackage[utf8]{inputenc} 
\usepackage[T1]{fontenc} 
\usepackage{comment}
\usepackage{algorithm}
\usepackage{algorithmic}
\usepackage{amsmath}
\usepackage{adjustbox}
\usepackage{subfig}
\usepackage{amsfonts} 
\usepackage{lipsum}
\usepackage[square,numbers]{natbib}
\usepackage{hyperref}
\usepackage{url}

\title{Fast 3D Volumetric Image Reconstruction from 2D MRI Slices by Parallel Processing}

\author{Somoballi Ghoshal$^{1*}$, Shremoyee Goswami$^{1}$, Amlan Chakrabarti$^{1}$, Susmita Sur-Kolay$^{2}$\\~\\
$^1$A. K. Choudhury School of Information Technology, University of Calcutta\\
$^2$Advanced Computing and Microelectronics Unit, Indian Statistical Institute\\~\\
$^*$somoballi@gmail.com}

\begin{document}
\maketitle

\begin{abstract}
Magnetic Resonance Imaging (MRI) is a technology for non-invasive imaging of anatomical features in detail. It can help in functional analysis of organs of a specimen but it is very costly.  In this work, methods for (i) virtual three-dimensional (3D) reconstruction from a single sequence of two-dimensional (2D) slices of MR images of a human spine  and brain along a single axis, and (ii) generation of missing inter-slice data are proposed. Our approach helps in preserving the edges, shape, size, as well as the internal tissue structures of the object being captured.  The sequence of original 2D slices along a single axis is divided into smaller equal sub-parts which are then reconstructed using edge preserved kriging interpolation to predict the missing slice information. In order to speed up the process of interpolation, we have used multiprocessing
by carrying out the initial interpolation  on  parallel cores.  From the 3D matrix thus formed, shearlet transform is applied to estimate the edges considering the 2D blocks along the $Z$ axis,  and to minimize the blurring effect using a proposed mean-median logic.  Finally, for visualization, the sub-matrices are merged into a final 3D matrix. Next, the newly formed 3D matrix is split up into voxels and marching cubes method is applied to get the approximate 3D image for viewing.  To the best of our knowledge it is a first of its kind approach based on kriging interpolation and multiprocessing for 3D reconstruction from 2D slices, and approximately 98.89\% accuracy  is achieved with respect to similarity metrics for image comparison. The time required for reconstruction has also been reduced by approximately 70\% with multiprocessing even for a large input data set compared to that with single core processing.
\end{abstract}


%
\maketitle

\section{Introduction}
%
%
%
%
Medical imaging captures the internal structure of a specimen. It can also help in the functional analysis of some of the organs or tissues  of the specimen \cite{1}.  There are several prevalent imaging technologies such as  medical ultrasonography or ultrasound, endoscopy, medical photography, elastography, tactile imaging, thermography, and then  radiological ones, namely X-ray radiography, computer aided tomography (CT), micro-computer tomography (micro-CT), magnetic resonance imaging (MRI), optical coherence tomography (OCT) \cite{1} . In this work, the focus is only on the most intensive one, i.e., MRI images.

\par
Magnetic resonance imaging (MRI) is  primarily used to form pictures of the anatomy and the physiological processes of the body in both health and disease \cite{3,10,2}. It is widely used for diagnosis of any brain and spine abnormality/disease  because it captures the tissue structure of the body most effectively. The diversity and complexity of lesion cells, particularly in functionally critical organs, make it very challenging to visualize a lesion in magnetic resonance imaging. This has led to 3D MRI for better visualization.

\par 3D MRI can be captured directly, but it comes at a colossal cost. The alternate most prominent approach for 3D MRI visualization is 3D reconstruction from 2D MRI slices. There are two types of 3D reconstruction techniques \cite{37}, namely, surface rendering and volume rendering. While surface rendering \cite{37} reconstructs to preserve the surface of the 3D object, volume rendering \cite{8} enables a volumetric visualization with the internal structure of the 3D object also reconstructed. Thus, when this 3D reconstructed data is sliced into 2D along a specified plane, we can also view the internal structures of the 3D object. 3D visualization has become salient for medical diagnosis \cite{31}, as it offers abundant and accurate information for medical experts. However, there are several challenges which need to be addressed. All the existing techniques mostly focus on preserving the shape of the object but not the internal structures. These also consider the information of all the three orthogonal planes $XY$, $YZ$, $ZX$. 

These techniques for 3D reconstruction of MRI of brain images have been exhibited in the literature \cite{5,6,9} but these fail for spine reconstruction. The reason is that in MRI of spine, the information of both hard and soft tissues are of equal importance, since the vertebrae as well as the bony spinal cord and inter vertebral discs are all present together. Thus, in the case of the spine, edge preservation while reconstruction is of utmost importance. This leads us to the goal of designing efficient 3D reconstruction for MRI of spine. In this work, the focus is on preserving the information of the internal tissues as well as the shape and size of the spine and on accelerating the speed for reconstruction. Moreover, all the existing interpolation techniques are deterministic and does not have any scope for error estimation while reconstruction, which can be overcome by the use of kriging interpolation. We have carried out 3D reconstruction using edge preservation by kriging interpolation, shearlet transform  along with multiprocessing. 
\par The elaborate explanation of our work  starts with motivation and related works in Sections II and III respectively followed by preliminary knowledge of the techniques used for interpolation, registration and quality assessment in Section IV. Our proposed methodology is described in Section V and our experimental results appear in Section VI. Section VII has a brief discussion along with conclusion.

\section{Motivation}
 Various existing techniques for 3D MRI scan capture a sequence of 2D image slices at equal intervals along  each of the three orthogonal axes and then the 3D image is generated \cite{1}. In case of CT scans, the slices along a single axis are captured and stacked to get the 3D. But this stacking only method results in loss of information in between the slices. In all types of medical imaging techniques, there is always a uniform slice gap ranging from 1 mm to 5 mm between two consecutive slices \cite{3}, hence a minimum of approximately 4-pixel information is missing between two slices which may result in erroneous internal structure reconstruction. The time taken for capturing  medical data for 3D reconstruction is approximately 45 mins \cite{9}, which means more energy is consumed and the computational complexity is also high. 

Our goal is to create efficient accurate 3D reconstruction and visualization of MRI from a sequence of 2D slices along a single axis, and also providing an user interface for the professionals to cut the reconstructed 3D image as needed with virtual scissors and to view any slice along any  plane. Thus, the scanning time can be reduced from 45 mins \cite{9} to close to the time taken to capture a single set of slices along a specified axis. The power consumption of a MRI machine \cite{38} to capture a patient is approximately 15 kW, which can be reduced  considerably with the reduction of time to capture the image. Since the time is proportional to cost, the huge cost of MRI especially in the developing countries,  can also be reduced and better health-care can be attained with cost, power consumption and exposure to radiation being minimized.

In \cite{12,39}, there are methods for 3D reconstruction and slicing of brain and spine respectively, but blurring along the edges do include 4\% error and these also fail for a large sized dataset. These works also deals with deterministic interpolation and spacial co-relation and error estimation is not incorporated. They also carry out 2D interpolation along an axis, thus all the voxels related to a voxel are never considered. In this work, we speed up the interpolation with error estimation while reconstruction based on spatial co-relation. The internal structures are preserved more accurately with the use of 3D interpolation for inter slice missing data reconstruction considering the information of all possible 64 neighbouring voxels. The error for reconstruction is approximately reduced to 1\% for real MR images. Our main contributions are proposing fast methods for:

\begin{itemize}

\item reconstructing a 3D volumetric image of a human spine or a brain from a sequence of 2D slices along a single axis, with preservation of internal structure by using edge preserving kriging interpolation on the voxels which in turn helps in shape and size restoration; 
\item dividing the image sequence into sub-sequences (8 here) and implementing the interpolation in parallel to attain a speed up of approximately 70\% ; 

\end{itemize}

To the best of our knowledge, this is the first of its kind method for 3D reconstruction  from a sequence of slices in one plane using parallel processing and for slicing this regenerated 3D as per user instructions to visualise the internal structure with a feature of 3D crop and zooming in.

\section{Related Works}

Several works for 3D reconstruction from 2D images in medical imaging domain \cite{4,31,14,28,33,35,26,44,45,46}  exist, but they fail to give good accurate
results with internal structure preservation very fast. While most methods suffice for CT images as the information is mostly of
hard tissue structure and outlines, but for MRI have lacuna with preservation of the internal tissue structure. 
The Marching cubes \cite{15} algorithm is widely accepted for
reconstructing a 3D surface from a given 3D image. For
approximating contours, it uses patterned cubes or iso-surfaces.
However, it requires certain techniques to reduce memory and time for reconstructing a surface from large volumetric data. The usual way to solve this problem \cite{4} is by diminishing the size of a volumetric image, but the quality of the surface of 3D reconstructed image becomes substandard if only sub-sampling is applied. Due to poor reconstruction by only volumetric sub-sampling, another method is proposed, which improves the quality of a surface reconstructed from the sampled volumetric data. It is based on a pipeline of Visualization Toolkit (VTK) \cite{22,21}. Their approach  consists of three major steps: preprocessing,
reconstructing and displaying. In \cite{25}, the authors have used kriging interpolation for reconstruction of inter-slice data but it gives a blurring effect not only along the boundary but along all the edges, thus making the reconstructed images noisy.  In \cite{23}, the authors have used tri-linear interpolation for reconstruction of MRI of brain and it gave better results than marching cube, but again fails in preserving the minute internal details.  In \cite{34}, the authors have used edge based interpolation for
correction of blurred and noisy edges in a 2D plane, but this work is inadequate for generating a large number of missing points in a 3D image. For 3D reconstruction by this technique applied on our real data, the accuracy after slicing is obtained to be 75.47\%. 

Several works for 3D reconstruction specifically for MRI \cite{5,6} also exist but they consider all three planes for reconstruction. In \cite{12}, the authors have used bilinear interpolation for reconstruction of 3D brain from 2D slices, along a single plane but in this case the edges are not properly preserved. Further, in \cite{39}, the authors have  reconstructed 3D image of human spine from a sequence of slice images along a single axis with 96\% accuracy by employing a combination of bilinear and bicubic algorithm, but the time complexity is very high even with a small number of inputs and a blurring effect along the boundary remains. It is unable to process a large sized dataset. 

In \cite{43}, the authors use deep learning for interpolating the missing data in between slices which gives good results for the trained set of data with no anomaly, but fails if there exist any deformity. Also, as every individual is unique,  generating the missing slices based on the information of the slices of some other individuals does not always give accurate results. Moreover, the structure of spine is always different in men, women and child and it also varies a lot  with height. None of the existing deep learning techniques have taken onto account all of these and hence fails in most cases and are not used or accepted by medical practitioners.

\par 
There are a few techniques for MRI reconstruction using multiple cores \cite{40,42} but these take into consideration the information for all three planes either while capturing the data (real time) or after capturing the full information along all three planes. Thus, we propose an advanced technique of 3D reconstruction from a single sequence of slices which takes information only from its preceding and succeeding slices to generate the missing information with edge preservation and also multiprocessing to speed up the process of 3D reconstruction.

\section{Methodology}

\begin{figure}[h!]
	\centering
	\includegraphics[width=8.2cm, height=12cm]{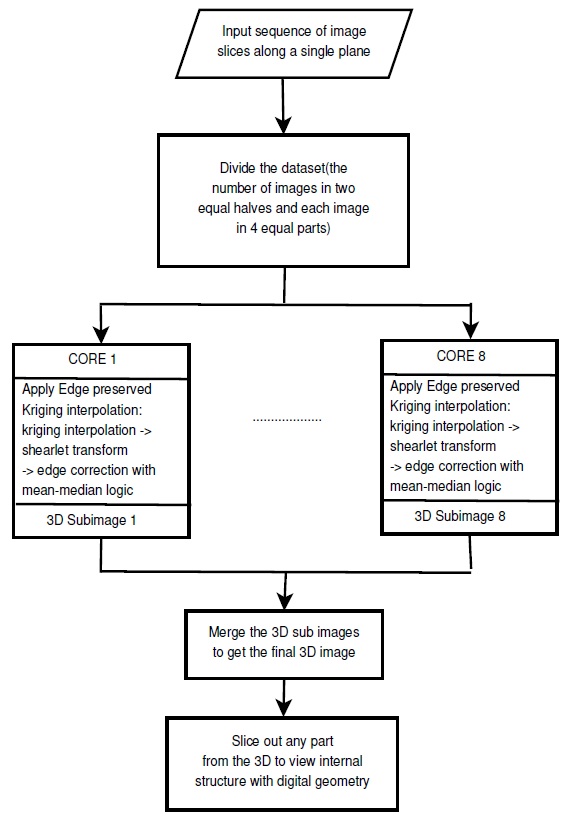}
	\hspace{.01 mm}
     \caption{Block diagram for 3D reconstruction from a single sequence of slices along one axis}	
     \label{blockdia}
\vspace{0.01 mm}
\end{figure}

Figure \ref{blockdia} shows the step-wise overview of our proposed approach. At first, the 2D slices along a single axis are denoised using shearlet transform \cite{24} (see Appendix~\ref{shear}). Next, we divide the data set into $N$ (a power of 2) equal parts. These $N$ sub sequences of 2D images are interpolated in parallel, using kriging interpolation \cite{25} (see Appendix~\ref{kriging}) to generate the missing slices in between two consecutive slices. Single instruction multiple data architecture (SIMD) is used to execute these $N$ blocks in parallel by multiprocessing on $N$ cores
to speed up the process of reconstruction.  From these 3D sub matrices, we then consider the 2D images along the $YZ$ plane and apply Shearlet transform. Edge detection is carried out as in \cite{29} and  the blur effect along the edges is minimized by a proposed mean-median logic.  
In order to get the final 3D matrix of the entire object we need to merge the 3D sub matrices that are obtained thus far. After joining the sub matrices, we break the 3D matrix into voxels and further apply the Marching cubes algorithm followed by smoothing \cite{32} for 3D reconstruction. We apply color map and activate rotation operation and then display the image on screen so that
the user can rotate the 3D image and view all sides of it as needed.  

Since we have generated the complete 3D volume data, we can easily cut and view  any specified required part of the object as required.

\subsection{3D reconstruction from 2D slices}
The proposed method for 3D reconstruction has three major components: parallel processing of the 3D reconstruction to gain speed,  edge preserved Kriging interpolation to predict the missing data in between slices, followed by merging the 3D reconstructed parts in parallel for visualization of the complete object.

\subsubsection{3D reconstruction with multiprocessing}
We assume that the slices registered with each other and are numbered in sequence.  The final 3D image to be reconstructed is represented
by a 3D matrix $M(i, j, k)$, which has a typical size of $m$ $\times$ $m$ $\times$ $(gn)$ as  the 2D slices are of dimension $m$ $\times$ $m$ and there are $n$ slices and $g$ is the number of pixels missing between slices. Initially, we split each image slice in $4$ equal parts, bisecting the image horizontally as well as vertically, thus each sub-image is of size $m/4$ $\times$ $m/4$ . We consider the leftmost corner of the original slice to be $(0, 0)$ and based on this assumption, the corresponding coordinate values of the split image with respect to the original slice are stored in an array $A$.  Now, for 3D reconstruction, we again divide the data set in $k$ equal parts, i.e., each sub set of images has $n/k$ slices in sequence.

For example, if we have a data set in which there are $50$ slices of size $1024 \times 1024$ and $k$ is 2, then we divide this input sequence into $8$ blocks with each block having $25$ slices of size  $512 \times 512$. Single instruction multiple data architecture (SIMD) is used to execute these $8$ blocks in parallel on 8 cores. The main part of the reconstruction procedure is the interpolation, hence by dividing the input set and applying multiprocessing on this yields an effective speedup. The input data is divided in such a way that they are mutually exclusive.

\begin{algorithm}
\begin{algorithmic}
\STATE \textbf{Input:} A sequence of $n$ 2D  slices along a plane $M$, $s_{gap}$ the inter-slice gap, integer $k$

\STATE\textbf{Output:} Set of 3D reconstructed sub images
\STATE \textbf{begin}	

  \STATE \hspace{0.1cm}\textit{Step 1:} Divide each image $M(i)$ into $4$ equal quadrants
\STATE \hspace{1.2cm} to get $4$ subsets of slices $M_1$ $\ldots$ $M_{4}$; 

\STATE\hspace{0.1cm}\textit{Step 2:} Divide each of the $4$ subsets into $4k$ subsets by  
\STATE \hspace{1.2cm} dividing the $n$ slices to get $k$ subsets  $M_1$ $\ldots$ $M_{4k}$ 
\STATE \hspace{1cm} of $n/k$ slices each;

\hspace{0.1cm} \textit{Step 3:} 
Execute  $R_i$ $=$ $Conc(M_i)$ on $4k$ parallel cores for  
\STATE \hspace{1.2cm} all the subsets $i=$ $1$ to $4k$; 
\STATE\hspace{1.2cm}/*partially fill up 3D matrices $R_i$, where $ i=$ $1$ to 
\STATE \hspace{1.2cm} $k$ with the available data $M_i$ along
\STATE \hspace{1.2cm}  the missing  axis for slice gap $s_{gap}$ */

\STATE \hspace{0.1cm}\textit{Step 4}
        Apply $F(i)[x,y,z]=$Edg\_prsvd\_Kriging$($R(i)$)$;
        
\STATE \hspace{0.1cm}\textit{Step 5} Store a set of $4k$ 3D sub images.
\STATE\textbf{end}
\end{algorithmic}
\caption{ Proposed Algorithm : 3D reconstruction from 2D MR image using multiprocessing}
\label{algo1}
\end{algorithm}

At first, for each block, a 3D matrix is created with the given data and each inter slice gap is padded with $NaN$. For each block, we calculate the maximum and minimum possible value, assuming that the measurement precision is correct upto 6 decimal points. Edge preserved Kriging interpolation is applied to calculate the value of the missing pixels. After this, we get 3D sub-images in each core which are to be joined together to get the final 3D volumetric image. Algorithm \ref{algo1} shows a step wise implementation of this approach.

\subsubsection{Edge preserved Kriging interpolation}

Initially, the 2D sub sequences of images are interpolated using kriging interpolation, in parallel, to generate the missing slices in between two consecutive slices.  Universal kriging \cite{25}, the widely used technique in spatial analysis and computer experiments, has been chosen for interpolation. It is applied in 3D domain considering all possible 64 neighbours  of a voxel. It gives better result than other state of the art interpolation technique as it computes the value for the unknown data point using a weighted linear sum of
known data values. The weights are chosen to minimize
the estimation error variance and to maintain unbiasedness. We get a 3D sub matrix with the interpolated slices that fill in the missing data. From this 3D matrix, we now consider the 2D images along the $YZ$ plane. Shearlet transform is again applied on these 2D images. We carry out multiprocessing and $8$ images are evaluated simultaneously to speed up the process. After applying shearlet transform, edge detection is carried out as in \cite{29}. We can easily get the orientation of each edge point $e$ from the shear coefficients \cite{29} as:

\begin{equation}
\theta_j(e) = \arg \max\limits_k|{SH(I)(j, k, e)}|.
\end{equation}

The blur effect incurred due to interpolation, is minimised by moving in clockwise direction and checking for $\theta_j$ of every edge point.  We consider a window of 16 edge pixels every time we traverse. If the direction of the edge points change frequently back and forth then we assume that there is noise. Then, we calculate $mean (\theta_j)$ and choose the points as edge with the mean value of $\theta_j$ corresponding to the earlier detected edge points. 

Suppose there is an edge orientation sequence  $\theta_1\theta_2\theta_1\theta_2\theta_3\theta_1\theta_2\theta_1\theta_1\theta_2\theta_1\theta_1\theta_2\theta_1\theta_2\theta_1$, then only the edge orientation for the initial and final pixel remain same and that for all the others changes to $mean(\theta_j)$, i.e., $\theta_1$ in this case,  and the edge points are changed accordingly. The pixel values adjacent to the newly selected edge pixel $e(x, y)$ are to be adjusted as follows: if the next selected pixel in clockwise direction is $e(x, y+1)$ then all the other pixels adjacent to $e(x, y)$ towards the left of it are assigned the median value of the pixels $e(x-1, y)$, $e(x-1, y+1)$, and $e(x-1, y-1)$, and the pixels adjacent to $e(x, y)$ towards the right of it are assigned the median value of the pixels $e(x+1, y)$, $e(x+1, y+1)$, and $e(x+1, y-1)$. Algorithm \ref{edgK} presents the steps of this approach.

\begin{algorithm}
\begin{algorithmic}
\STATE \textbf{Input:} Reconstructed 2D sub-images $R(i,j,1)$ extracted along $z$ axis
\STATE\textbf{Output:}  Edge preserved 2D sub-images $R(i,j,1)$
\STATE \textbf{begin}	
             \STATE \hspace{0.1cm}\textit{Step 1:} Apply shearlet transform;
             
            \STATE \hspace{0.1cm}\textit{Step 2:} Find the edges and the orientation $\theta$ of each edge \STATE \hspace{1.2cm} pixel;
            
             \STATE \hspace{0.1cm}\textit{Step 3:} Select an arbitrary edge pixel and move in 
             \STATE \hspace{1.2cm} clockwise direction for a window of 16 pixels;
             \STATE \hspace{1.2cm} (size of window may be chosen as per the dataset) 
              
              \STATE \hspace{0.1cm}\textit{Step 4:} Find the mean $\theta$ of these 16 pixels and select the \STATE \hspace{1.2cm} new edge pixel with $\theta$ orientation from the 8
              \STATE \hspace{1.2cm}  neighbours of the previous edge pixel if 
              \STATE \hspace{1.2cm} orientation of edge pixel is not $\theta$;
             
            \STATE \hspace{0.1cm}\textit{Step 5:} Adjust the 8 neighbour pixels of the newly found 
            \STATE \hspace{1.2cm} edge pixel $(x,y)$ by replacing 3 adjacent pixels to 
            \STATE \hspace{1.2cm} the left of it with their median, and the 3 adjacent 
            \STATE \hspace{1.2cm} pixels to the right of it with their median, 
            i.e,
              \STATE
           \STATE \hspace{1.2cm}$e(x$-$1, y)$ = $e(x$-$1, y$+$1)$ = $e(x$-$1, y$-$1)$ = \\  
           \STATE \hspace{1.2cm}$ median(e(x$-$1, y), e(x$-$1, y$+$1), e(x$-$1, y$-$1))$;
          \STATE
           \STATE \hspace{1.2cm} $e(x$+$1, y)$ = $e(x$+$1, y$+$1)$ = $e(x$+$1, y$-$1)$ = \\  
           \STATE \hspace{1.2cm}$ median(e(x$+$1, y), e(x$+$1, y$+$1), e(x$+$1, y$-$1))$.
          
\STATE\textbf{end}
					
\end{algorithmic}
\caption{ Edge preserved Kriging interpolation()}
\label{edgK}
\end{algorithm}

\subsection{3D visualization of the complete object}

For 3D visualization, we need to merge the 3D sub matrices by incorporating all the 3D sub matrices into a single 3D matrix based on their reference position. While dividing the data in eight equal parts we have stored the matrix indices to which they will correspond in the final image. Thus, by simply filling in the data form the sub matrices based on the matrix index we get the final 3D image.
Then,  for better viewing, we break the 3D matrix in voxels and further apply the Marching cubes algorithm followed by smoothing \cite{32} for 3D
reconstruction. We apply color map and activate rotation operation and display the image on screen so that
the user can rotate the 3D image and view all sides of it as needed.

\subsection{3D visualization of a particular section of the object}

The reconstructed volume data, $\mathcal V$ = $F$, is simply a set of voxels (i.e., 3-cells \cite{79}), and it can easily be cut and viewed using digital planes. 
The user needs to specify as input the view-direction and view-depth. 
Accordingly, the real plane $P$ with the corresponding orientation and depth in the local coordinate system is considered. 
Then, its corresponding thinnest digital model 
${\mathbb D}(P)$\footnote{referred to as {\em $2$-minimal digital plane} 
in the literature of digital geometry \cite{79}} is taken up, 
and its intersection with $\mathcal V$ is computed using elementary set-theoretic operations. 
As ${\mathbb D}(P)$ is 2-minimal, it partitions $\mathcal V$ into two well-defined components that are guaranteed to be 2-separable from each other; 
that is, there does not exist any 2-connected path of voxels from one component to the other. 
If the user  wants to isolate, cut, or process a block out of the 3D volume, then he/she needs to specify the cut-planes accordingly, as given in Algorithm~\ref{algo2}. 
As it involves thinnest digital planes, the operations with intersecting, cutting, joining, and related procedures including viewing are fast and efficient.

\begin{algorithm}
\begin{algorithmic}
\STATE \textbf{Input:} Reconstructed 3D sub-images, $R_i$, array of boundary coordinated, $C$, user specified coordinates of the bounded area to be extracted and displayed, $UC$
\STATE\textbf{Output:}  3D image for visualization, $PR$, of the required area
\STATE \textbf{begin}	
            
 \STATE \hspace{0.1cm}\textit{Step 1:} Search for the specified coordinates,$UC$ in $C$, 
   \STATE \hspace{1.2cm} Choose all $R_i$ that contains $UC$
     \STATE \hspace{1.2cm} Merge the thus found $R_i$ to get $PR$;
  
 \STATE \hspace{0.1cm}\textit{Step 2:} 
   if size($PR$) $>$  ($512 \times 512 \times 40$) then
 \STATE \hspace{1.2cm} Resize the  $PR$ to ($512 \times 512 \times 40$) using seam 
  \STATE \hspace{1.2cm} carving technique;
             
 \STATE \hspace{0.1cm}\textit{Step 3:} Fill $PR$ with $0$ for the area that lies outside the   
 \STATE \hspace{1.2cm} specified bounded region;
             
 \STATE \hspace{0.1cm} \textit{Step 4:}  Apply marching cube on $PR$; 

\STATE \hspace{0.1cm} \textit{Step 5:} R = $Colormap$(PR);	/*MATLAB function*/		

\STATE \hspace{0.1cm} \textit{Step 6:} $Rotate\_para$(R); /*MATLAB function*/

\STATE \hspace{0.1cm} \textit{Step 7:} $Display$(R). /*MATLAB function*/

\STATE\textbf{end}

\end{algorithmic}
\caption{ Proposed Algorithm : 3D visualization of required specified area}
\label{algo2}
\end{algorithm}

\section{Experimental Analysis}

We validated our fast 3D volumetric image reconstruction method on 24 real life T2-MRI data ($512\times512$ pixels) of human spine  with an inter slice gap of $3$ to $5 mm$, and 30 real life T2-MRI data ($512\times512$ pixels) of human brain  collected from Bangur Institute of Neurosciences, S.S.K.M, Kolkata and brain MRI data set of python.
 
\subsection{Multiprocessing}
Multiprocessing was carried out on a 32 core CPU with 64 GB memory. The edge preserved kriging interpolation for 3D reconstruction form 2d slices was executed in parallel for all submatrices. In addition to employing multiprocessing with 8 cores, we have also experimented  with 16 cores by breaking the data set into further smaller parts. But in this case the overhead time of joining the 3D submatrices is high, thus no evident gain in time is obtained compared to that with 8 cores. We also performed the experiment with 4 cores by  dividing each slice only into 4 parts but not halving the sequence of 2D slices. In this case, the time required to execute the interpolation for large sized data was approximately 2 mins 23 seconds which is higher than that for using 8 cores. Thus, we decided to continue with 8 cores to execute and speed up the reconstruction for our entire dataset. Table \ref{tab_time} in  shows the comparative time and overhead required with respect to the number of cores. The times required are an approximate value with an average image size $1024\times 1024$, and input set of 44 slices. The overhead time in this case is considered to be the time required to split the input dataset, the transfer time of each input set from CPU to each core, transfer time of each result from each core to CPU, time required to join all the sub matrices. We have tested our approach by decreasing or increasing the number of cores by a factor of 2 for the ease of splitting the data.

\begin{table}
\centering
\caption{Comparative study of number of cores with respect to execution time and overhead-time}
\label{tab_time}
 \begin{tabular}{|c|c|c|}
\hline
&&\\
No. of Cores  &Reconstruction time   &Overhead time\\
&&\\
\hline
&&\\
1&3 mins&5seconds\\
&&\\
\hline
&&\\
4&1 mins  8 seconds& 15 seconds\\
&&\\
\hline
&&\\
8&34 seconds&20 seconds\\
&&\\
\hline
&&\\
16&20 seconds&34 seconds\\
&&\\
\hline
 \end{tabular} 
 \end{table}

\begin{figure}[]
\centering
\subfloat[Original image slice]{\label{fig:ls}\includegraphics[width=2.5cm,height=4cm]{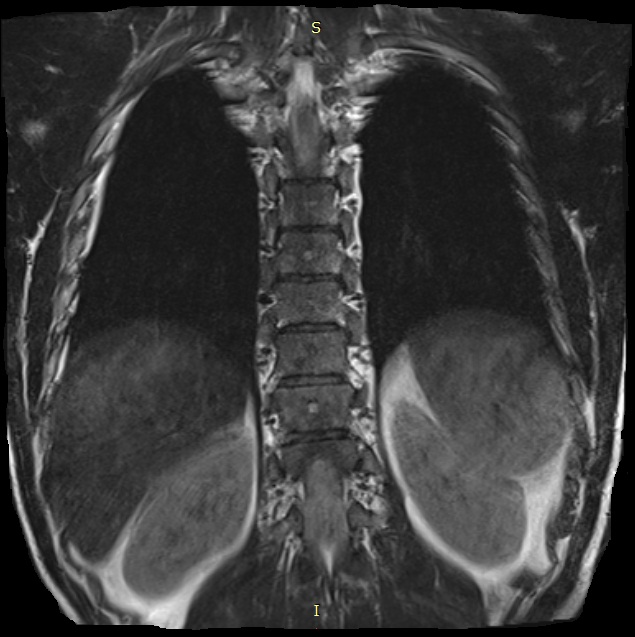}}
         \hspace{1 cm}
         \vspace{0.2mm}
	\subfloat[Image slice divided into 4 equal parts ]{\label{fig:split}\includegraphics[width=2.5cm,height=4cm]{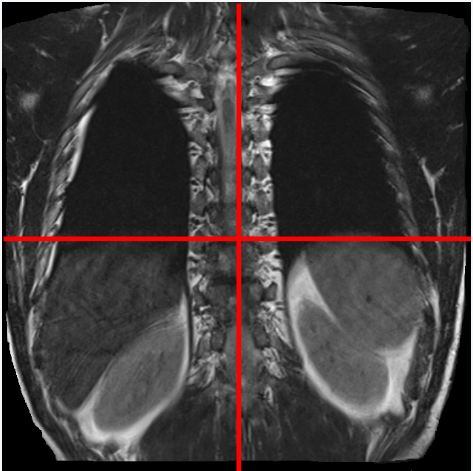}}
         \hspace{2 cm}
         \vspace{0.2mm}
	\subfloat[Sub-images of the slice]{\label{fig:splitted4}\includegraphics[width=7cm,height=2.5cm]{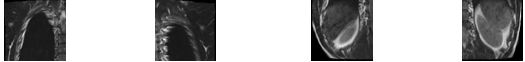}}
	\caption{Splitting a 2D image slice for a human spine}	
	\label{MRspinesplit}
\end{figure}

\begin{figure}
	\centering
	\includegraphics[width=.48\textwidth]{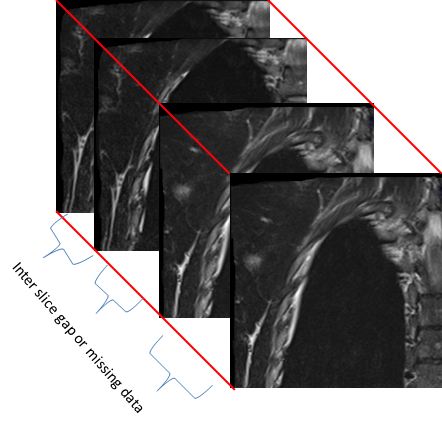}
	\hspace{.01 mm}
     \caption{Sequence of original 3D sub-images with missing data for a human spine}
     \label{stack}
\vspace{0.01 mm}
\end{figure}

\begin{figure}[]
\centering
	\subfloat[]{\label{fig:ri1}\includegraphics[width=4cm, height=4cm]{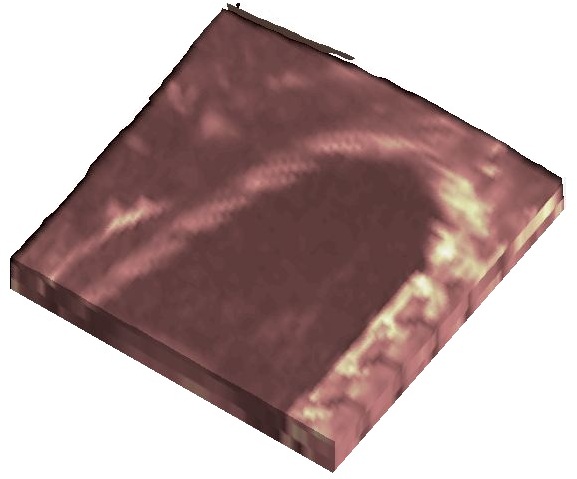}}
         \hspace{.01 mm}
	\subfloat[]{\label{fig:ri2}\includegraphics[width=4cm,height=4cm]{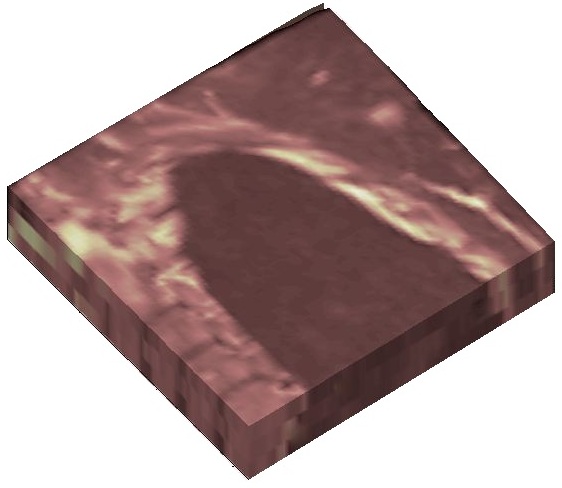}}
\hspace{.01 mm}
	\subfloat[]{\label{fig:ri3}\includegraphics[width=4cm, height=4cm]{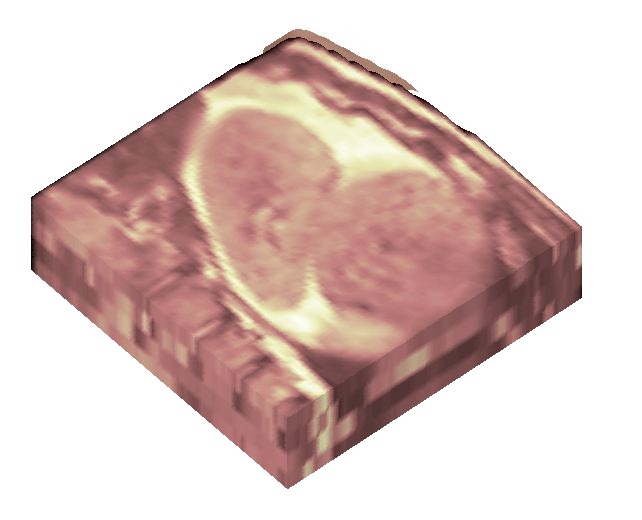}}
	\hspace{.01 mm}
		\subfloat[]{\label{fig:ri4}\includegraphics[width=4cm, height=4cm]{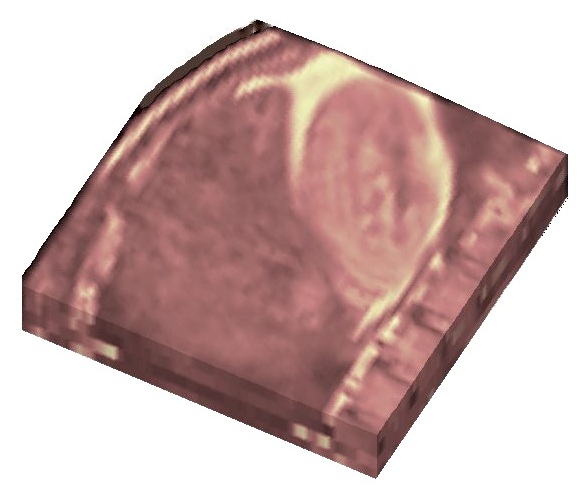}}
		\hspace{.01 mm}
			\subfloat[]{\label{fig:ri5}\includegraphics[width=4cm, height=4cm]{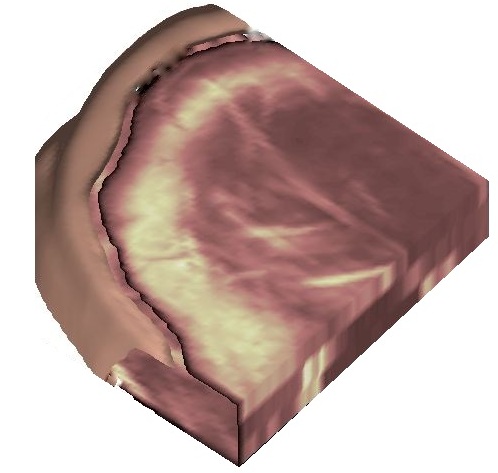}}
			\hspace{.01 mm}
			\subfloat[]{\label{fig:ri6}\includegraphics[width=4cm, height=4cm]{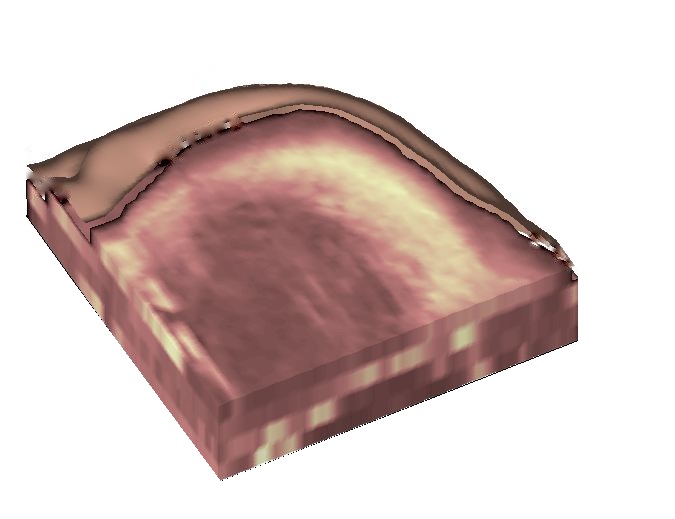}}
			\hspace{.01 mm}
			\subfloat[]{\label{fig:ri7}\includegraphics[width=4cm, height=4cm]{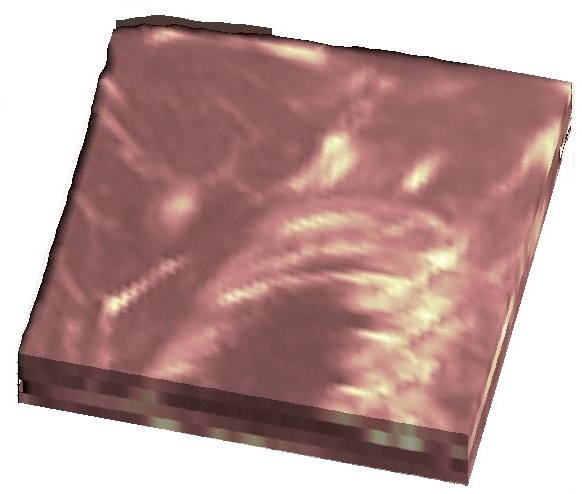}}
			\hspace{.01 mm}
			\subfloat[]{\label{fig:ri8}\includegraphics[width=4cm, height=4cm]{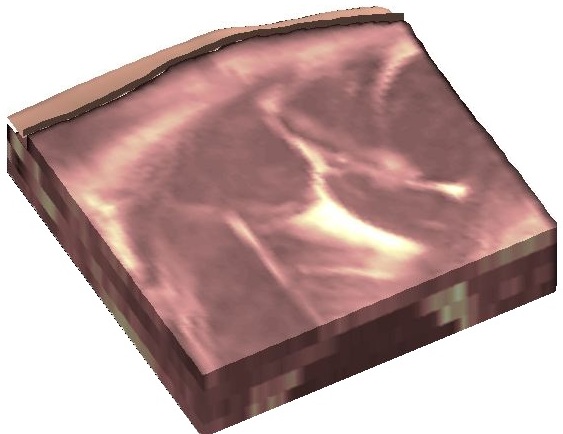}}
			\hspace{.01 mm}
	\caption{3D reconstructed sub-images for a human spine}	
	\label{3Dsub}
\vspace{0.2mm}
\end{figure}

\begin{figure}
	\centering
\subfloat[]{\includegraphics[width=3cm, height=5cm]{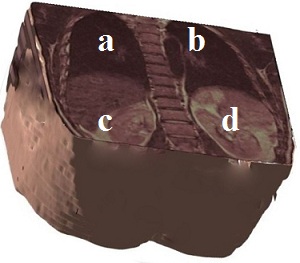}}
	\hspace{.02 mm}
	\subfloat[]	{\includegraphics[width=3cm, height=5cm]{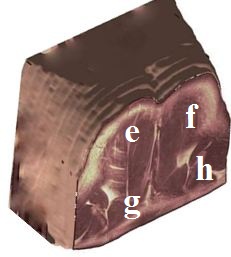}}
	
     \caption{Complete 3D visualization for a human spine after joining the eight 3D sub-images: (a) Anterior view, (b) Posterior view  [a,b,c,d,e,f,g,h marked on this figure corresponds to the sub-images in Figure \ref{3Dsub}] }	
     \label{3Dspine}
\vspace{0.01 mm}
\end{figure}

\begin{figure}[]
\centering
	\subfloat[Captured slice \#90 along the coronal plane for the slice sequence of Figure 1]{\includegraphics[width=3.5cm,height=4.6cm]{lumbarspineslice90.jpg}}
         \hspace{.4 mm}
         \vspace{0.2mm}
	\subfloat[Reconstructed slice 90 along the coronal plane for slice sequence of Figure \ref{3Dspine} ]{\label{fig:ri}\includegraphics[width=3.5cm,height=4.6cm]{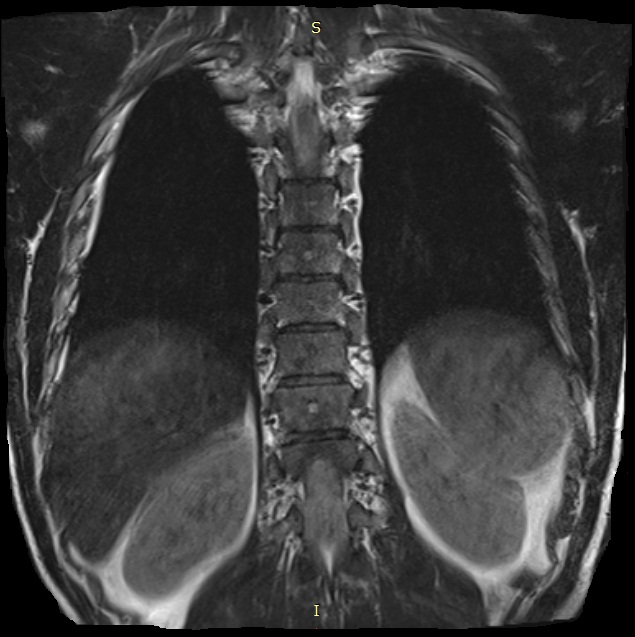}}
\hspace{.01 mm}
	\caption{Slice \# 90 for a human spine along the coronal plane as captured and after reconstruction}
	\label{brlumbar}
\vspace{0.2mm}
\end{figure}
\subsection{Our results for human spine and brain}
Initially,  the  $2D$ slices are split in 4 sub parts as shown in Figure \ref{MRspinesplit} for a human spine, and the data set in divided in two parts. We use single instruction multiple data architecture using 8 logical cores. In parallel, for each block of sub-image, a $3D$ matrix is created which is filled with the corresponding data leaving the specified slice gap in between slices as specified for each set as shown in Figure \ref{stack}. Then edge preserved  kriging interpolation is used to generate the 3d subimages. If we want to visualize these sub-images, then we can apply marching cube with color map and visualize the images as shown in Figure \ref{3Dsub} and the complete 3D image for visualisation of full spine is as in Figure \ref{3Dspine}. 

\begin{figure}[]
\centering
\subfloat[Input Sequence]{\label{fig:ri1}\includegraphics[width=7cm,height=7cm]{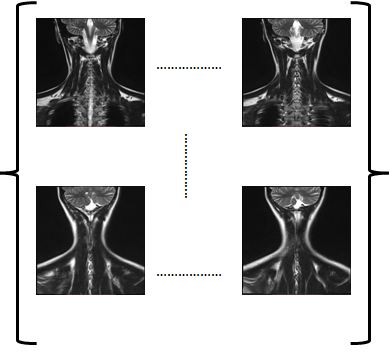}}
         \hspace{2 cm}
         \vspace{0.2mm}
	\subfloat[3D Reconstructed Image]{\label{fig:ri3d}\includegraphics[width=5cm,height=5cm]{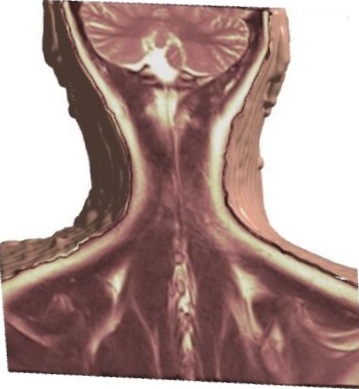}}
         \hspace{2 cm}
         \vspace{0.2mm}
	\subfloat[Axial Slice]{\label{fig:ri23}\includegraphics[width=2.5cm,height=3.5cm]{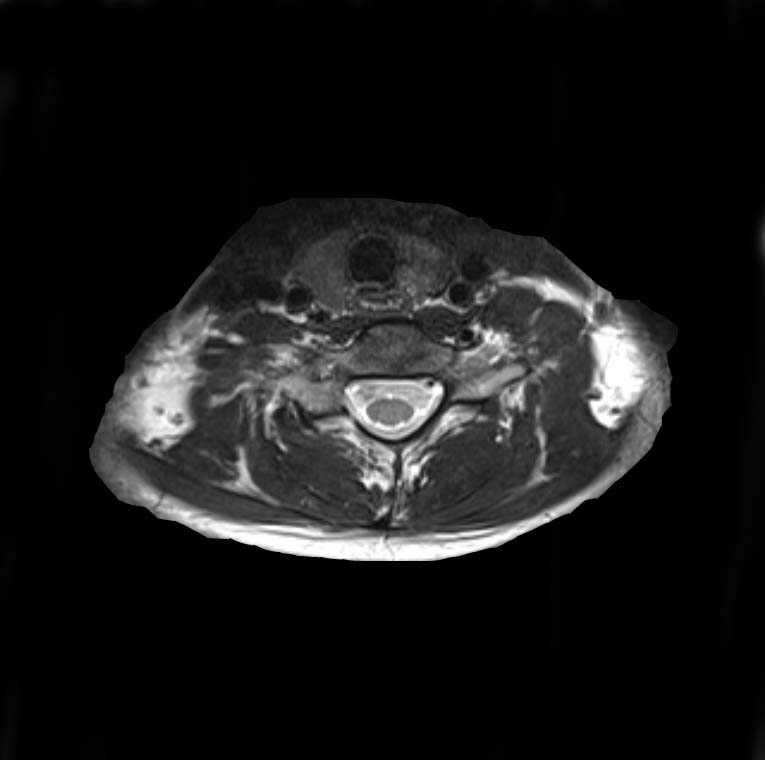}}
\hspace{.01 mm}
	\subfloat[Sagittal Slice]{\label{fig:ri7hcs}\includegraphics[width=2.5cm, height=3.5cm]{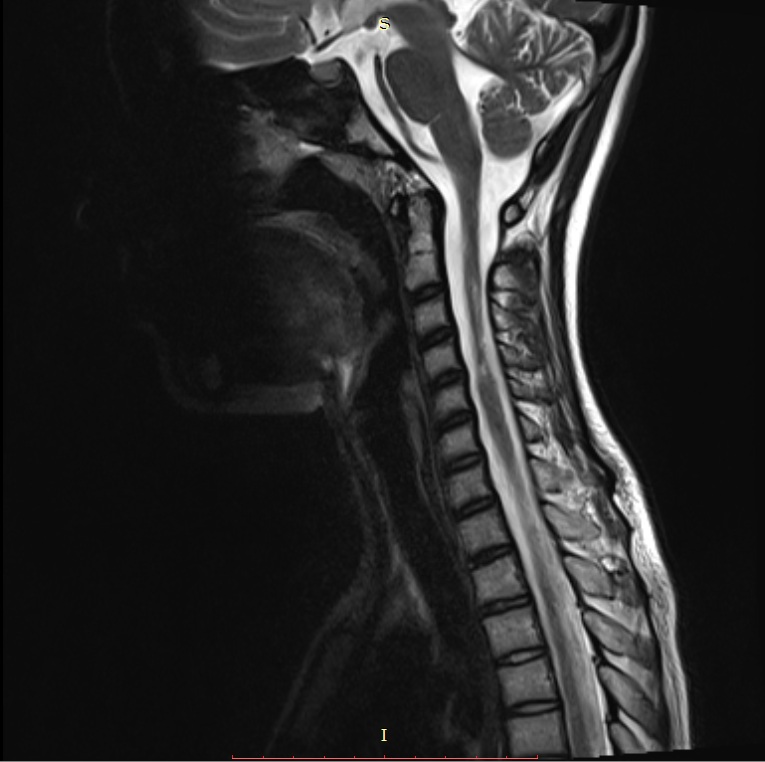}}
	\hspace{.01 mm}
	\subfloat[Coronal Slice]{\label{fig:ri8hcc}\includegraphics[width=2.5cm, height=3.5cm]{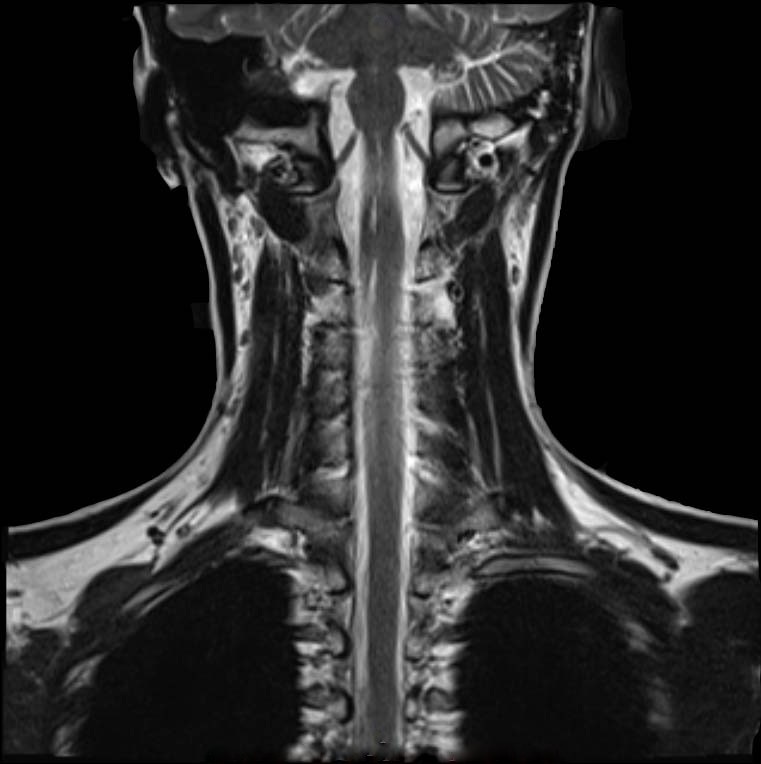}}
	\caption{3D Reconstruction of MRI for a human spine and slicing it along the three orthogonal planes as specified}	
	\label{MRspine}
\end{figure}

\begin{figure}[]
\centering
	\subfloat[Captured slice \#63 along the saggital plane for Slice sequence of Figure \ref{fig:ri1} ]{\includegraphics[width=3.5cm,height=4.6cm]{7hcs.jpg}}
         \hspace{.4 mm}
         \vspace{0.2mm}
	\subfloat[3D Reconstructed slice \#63 along the saggital plane  for slice sequence of Figure \ref{fig:ri3d} ]{\label{fig:ri}\includegraphics[width=3.5cm,height=4.6cm]{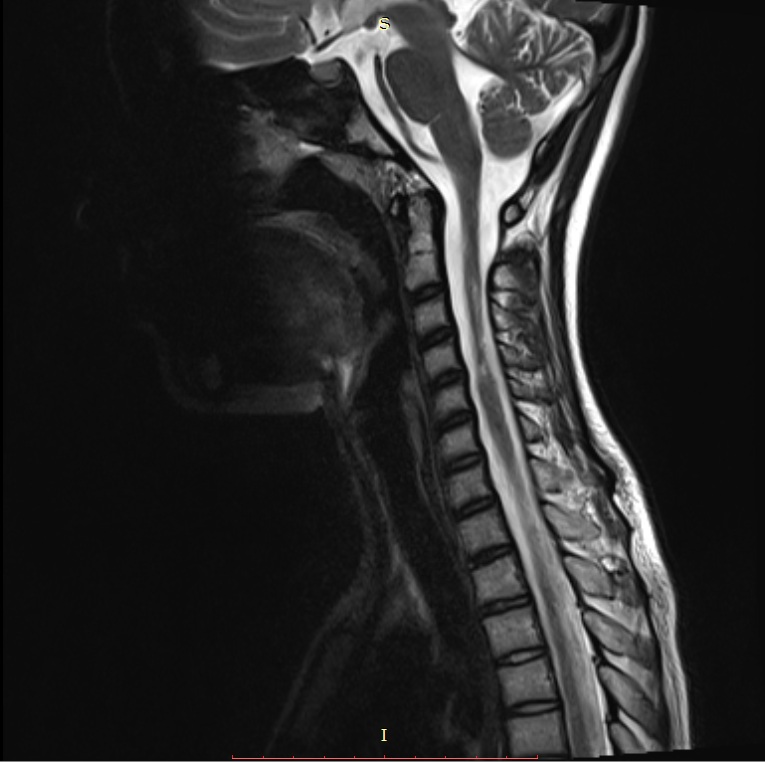}}
\hspace{.01 mm}
	\caption{Slice \#63 for a human spine along the saggital plane  for slice sequence of Figure \ref{fig:ri3d} while capture vs after reconstruction}
	\label{brsag}
\vspace{0.2mm}
\end{figure}

\begin{figure}[]
\centering
	\subfloat[Input Sequence of slices with slice gap of 3mm ]{\label{fig:rifull}\includegraphics[width=7cm, height=6cm]{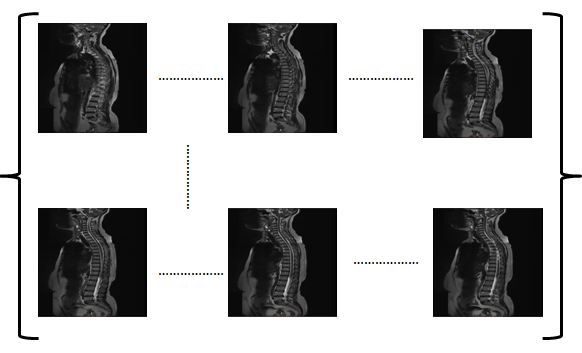}}
         \hspace{.01 mm}
         \vspace{0.2mm}
	\subfloat[Reconstructed 3D image]{\label{fig:ri3dfull}\includegraphics[width=5cm,height=6cm]{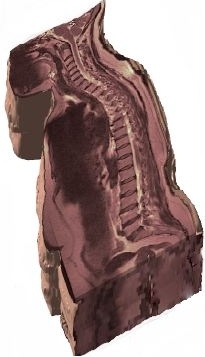}}
\hspace{.01 mm}
	\caption{3D reconstruction from MRI image sequence of a human full spine}	
	\label{brfull}
\vspace{0.2mm}
\end{figure}

\begin{figure}[]
\centering
	\subfloat[Captured slice \#148 along the axial plane for slice sequence of Figure \ref{fig:rifull}]{\includegraphics[width=3cm,height=2cm]{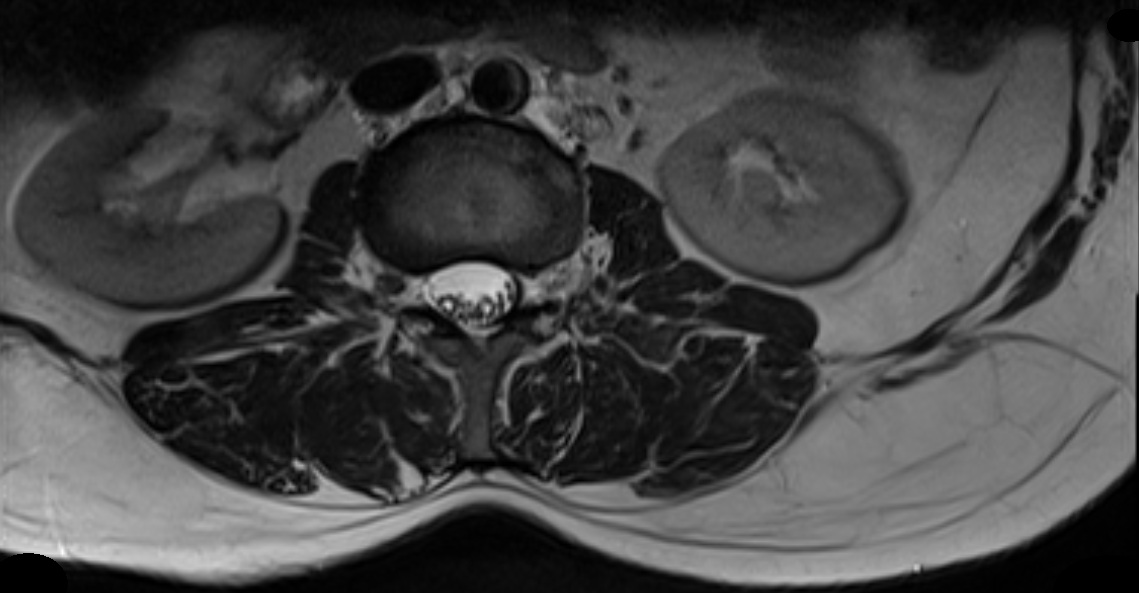}}
         \hspace{4 mm}
         \vspace{0.2mm}
	\subfloat[Reconstructed slice \#148 along the axial plane for slice sequence of Figure \ref{fig:ri3dfull} ]{\label{fig:ri48}\includegraphics[width=3cm,height=2cm]{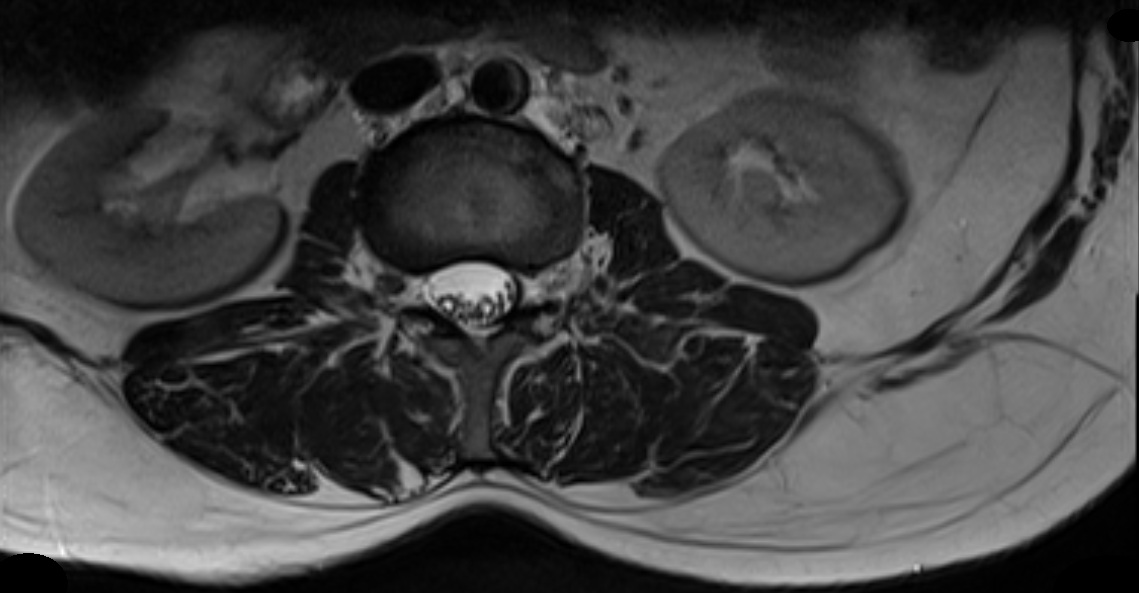}}
\hspace{.01 mm}
	\caption{Slice \#148 for a human spine along the axial plane for slice sequence of Figure \ref{fig:ri3dfull} while capture vs after reconstruction }
	\label{brax}
\vspace{0.2mm}
\end{figure}

\begin{figure}[]
\centering
	\subfloat[Input Sequence of slices with slice gap of 5mm  ]{\label{fig:ri}\includegraphics[width=7cm, height=6cm]{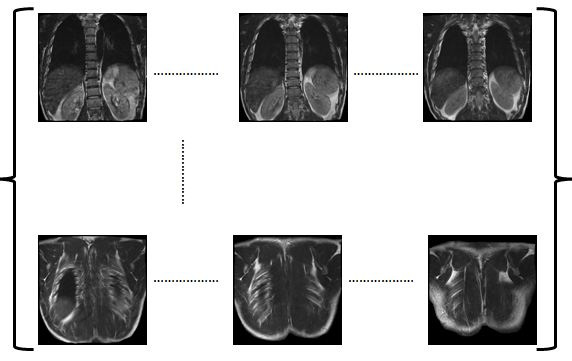}}
         \hspace{.01 mm}
         \vspace{0.2mm}
	\subfloat[Reconstructed 3D]{\label{fig:ri}\includegraphics[width=5cm,height=4cm]{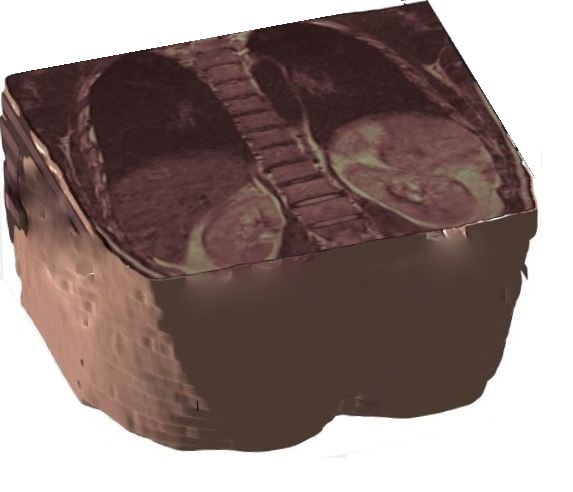}}
\hspace{.01 mm}
	\caption{3D reconstruction from MRI image sequence of human lumbar spine}	
	\label{brl}
\vspace{0.2mm}
\end{figure}
\begin{figure}[]
\centering
	\subfloat[Input Sequence of slices ]{\label{fig:ribraininput}\includegraphics[width=7cm, height=5cm]{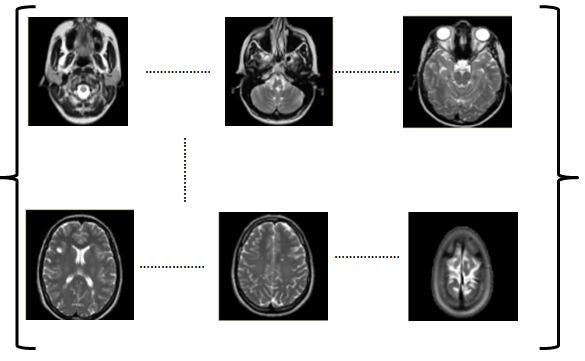}}
         \hspace{.01 mm}
         \vspace{0.2mm}
	\subfloat[Reconstructed 3D]{\label{fig:ribrain3d}\includegraphics[width=4cm,height=5.7cm]{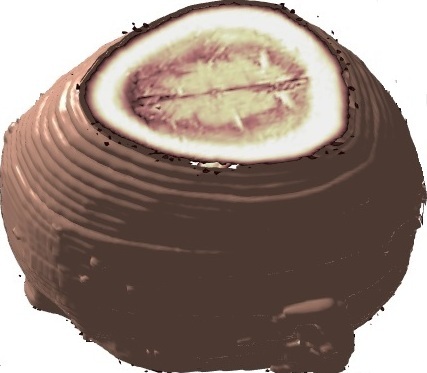}}
\hspace{.01 mm}
	\caption{3D reconstruction from MRI image sequence of human brain}	
	\label{mribra}
\vspace{0.2mm}
\end{figure}
\begin{figure}[]
\centering
	\subfloat[Captured slice \#40 along the saggital plane for Slice sequence of Figure \ref{fig:ribraininput}]{\includegraphics[width=3cm,height=2cm]{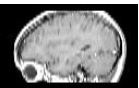}}
         \hspace{4 mm}
         \vspace{0.2mm}
	\subfloat[Reconstructed slice \#40  along the saggital plane, given the Slice sequence of Figure \ref{fig:ribrain3d} ]{\label{fig:ri}\includegraphics[width=3cm,height=2cm]{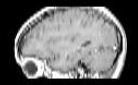}}
\hspace{.01 mm}
	\caption{Slice \#40 for a human brain along the saggital plane given the Slice sequence of Figure \ref{fig:ribrain3d} while capture vs after reconstruction}
	\label{brcompare}
\vspace{0.2mm}
\end{figure}

 In Figure \ref{MRspine} an example of 3D volumetric reconstruction, visualization as well as slicing are depicted. Figures \ref{brfull}, \ref{brl}, \ref{mribra} provide the results of our 3D reconstruction  from a sequence of 2D slices for a full spine, a lumbar spine and a brain respectively. We compared the results after slicing with the available ground truth data (Figures \ref{brlumbar}, \ref{brsag}, \ref{brax}, \ref{brcompare} for human lumbar along coronal, saggital, axial planes and human brain along sagittal plane respectively) based on mutual information, entropy, root mean square error and  structural similarity index (see Appendix~\ref{sm}). The average time taken by our $3D$ reconstruction fo human spine is 54 seconds, depending on the size of the input data set. The time taken for slicing is a fraction of a second. The average accuracy percentage of our method, as shown in Table \ref{tab:1},  is calculated as :

\begin{equation}
A_{imagetype}{\%}=Avg(A_{imagetype}^{ED}{\%},A_{imagetype}^{{MI}}{\%},A_{imagetype}^{{RMSE}}{\%},A_{imagetype}^{{SSIM}}{\%})
\end{equation}
where,
\begin{equation*}
A_{imagetype}^{ED}{\%}=\dfrac{1}{n}[\sum_{i=1}^{n}[1-[\dfrac{1}{m}\sum_{gap=1}^{m}\dfrac{\sum_{all\_slices,all\_gaps}ED(slice)}{total\_{slices}}]]]*100
\end{equation*}

\begin{equation*}
A_{imagetype}^{{MI}}{\%}=\dfrac{1}{n}[\sum_{i=1}^{n}[1-[\dfrac{1}{m}\sum_{gap=1}^{m}\dfrac{\sum_{all\_slices,all\_gaps}(MI(G,G)-MI(G,slice))}{total\_{slices}}]]]*100
\end{equation*}

\begin{equation*}
A_{imagetype}^{{RMSE}}{\%}=\dfrac{1}{n}[\sum_{i=1}^{n}[1-[\dfrac{1}{m}\sum_{gap=1}^{m}\dfrac{\sum_{all\_slices,all\_gaps}(RMSE(G,slice))}{total\_{slices}}]]]*100
\end{equation*}

\begin{equation*}
A_{imagetype}^{{SSIM}}{\%}=\dfrac{1}{n}[\sum_{i=1}^{n}[1-[\dfrac{1}{m}\sum_{gap=1}^{m}\dfrac{\sum_{all\_slices,all\_gaps}(SSIM(G,slice))}{total\_{slices}}]]]*100
\end{equation*}
and the average time is calculated as :
\begin{equation*}
\centering
T_{imagetype}=\dfrac{1}{s}[\sum_{i=1}^{s} time(t)]]
\end{equation*}
where $s$ is the number of  set of images of $imagetype$,
$n$ the number of samples of each type. Entropy difference ($ED$) is the percentage of difference in entropy between the sliced image and the original captured slice. Mutual information $MI$($G$, slice) is the  mutual information between the original ground truth image and the sliced image and $MI(G,G)$ is the mutual information if the original image is the sliced image. Root mean square error (RMSE) is computed between the ground truth image and the reconstructed image. Structural similarity index measure (SSIM) is the measure of structural similarity between the reconstructed image and the ground truth image. Table \ref{tab:1} shows the average mutual information,entropy difference,root mean square error and structural similarity index measure along each plane sliced after reconstruction, using our proposed method in comparison to the original data as shown in Figures \ref{brlumbar}, \ref{brsag}, \ref{brax} \ref{brcompare}. The average accuracy of the slices after applying our method along all 3 sequence of slices is $98.86\%$. Since the original data had a slice gap of either 5mm, 3mm or 1mm, these slices could be exactly matched with the original data set. The average accuracy of the slices generated after reconstruction compared to the original slices is 98.86\%. 

Table \ref{tab:3} gives a comparison  of the existing works for slice interpolation  with our proposed technique based on average accuracy, and  our algorithm outperforms the other methods.

\begin{table}
\centering
\caption{Average Mutual Information, Entropy Difference, Root Mean Square and Structural Similarity Index of slices along all three axes for cervical, lumbar and full spine dataset}
\label{tab:1}
 \begin{tabular}{|c|c|c|c|c|c|c|}
\hline
Sample  &Inter-slice  &Avg&Avg  &Avg  &Avg \\
&gap& MI&$EN\_D$&RMSE&SSIM\\
\hline
 &1mm &5.768 &0.001 &0.0001 &0.992\\
Brain &2mm&5.77 &0.001&0.0001&0.992\\
MRI&3mm&5.766&0.001&0.0001&0.99\\
&4mm &5.768&0.001&0.0001&0.99\\
&5mm &5.769 &0.001&0.0001&0.99\\
\hline
 &1mm &6.411 &0.001 &0.0001 &0.989\\
Spine &2mm&6.410 &0.001&0.0001&0.989\\
MRI&3mm&6.411&0.001&0.0001&0.989\\
&4mm &6.410&0.001&0.0001&0.988\\
&5mm &6.411 &0.001&0.0001&0.988\\
\hline
 \end{tabular} 
 \end{table}

\begin{table}
\centering
\caption{Comparison of accuracy of interslice reconstruction on the real data sets for human brain and spine that were available}
\label{tab:3}
 \begin{tabular}{|c|c|c|}
\hline
&&\\
Type of image &Algorithm  &Accuracy \%\\
&&\\
\hline
&&\\
&3plane method \cite{22}&98.8\% \\
&&\\

&&\\
&Kriging interpolation \cite{25}&85\%\\
&&\\

&&\\
Brain & Deep learning \cite{43}&94\% \\
&&\\

&&\\
&Bilinear \cite{12} &95\%\\
&&\\

&&\\
&Bilinear Bicubic Combined \cite{39}\ &97.86\%\\
&&\\
&&\\
&Proposed method&$\bold{99\%}$\\
&&\\
\hline
&&\\
&3plane method \cite{22}&98.8\%\\
&&\\

&&\\
&Kriging interpolation \cite{25}&80\%\\
&&\\

&&\\
Spine &Deep learning \cite{43}&90\%\\
&&\\

&&\\
&Bilinear \cite{12} &90\%\\
&&\\

&&\\
&Bilinear Bicubic Combined \cite{39}\ &96\%\\
&&\\

&&\\
&Proposed method&$\bold{98.9\%}$\\
&&\\
\hline
 \end{tabular} 
 \end{table}

%

 \subsection{User Interface}
\begin{figure*}
	\centering
\subfloat[Initial User Interface]{\label{fig:usr1}\includegraphics[width=7cm,height=5cm]{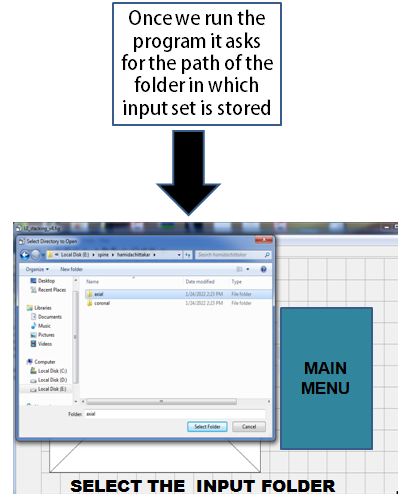}}
         \hspace{1 cm}
         \vspace{0.2mm}
	\subfloat[3D output with interface to view slices/a portion of the image ]{\label{fig:usr3}\includegraphics[width=7cm,height=5cm]{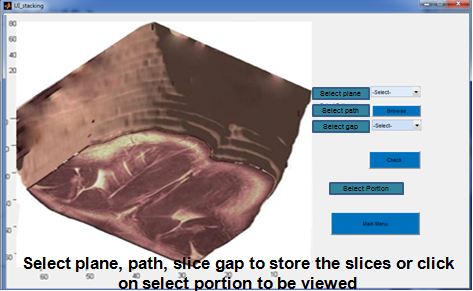}}
         \hspace{2 cm}
         \vspace{0.2mm}
	\subfloat[Interface after selecting a portion by user]{\label{fig:splitted4}\includegraphics[width=7cm,height=5cm]{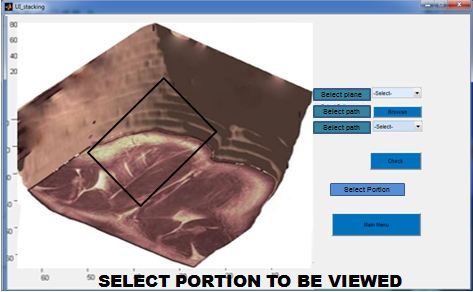}}
	 \hspace{2 cm}
         \vspace{0.2mm}
	\subfloat[Output after selecting a portion]{\label{fig:splitted4}\includegraphics[width=7cm,height=5cm]{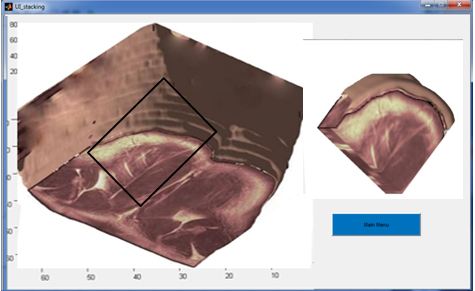}}	
	 \caption{User Interface}
	\label{usr}
\end{figure*}

We have designed an user interface as shown in Figure \ref{usr}, with which an user can generate a 3D view  from a set of 2D MRI slices along a single plane and slice out this 3D as per specifications along any plane or the user can cut out a portion of the thus formed 3D using virtual scissors.


\section{Concluding Remarks}

We have proposed an efficient multiprocessing based method for 3D volumetric reconstruction. We have carried out our experiments on  real life human  brain and spine  MRI collected from Bangur Institute of Neurosciences, Kolkata. For the purpose of reconstruction from large volume of data, we  divided  the data from its original size into $4$ equal parts and further divided the data set in two parts, thus ensuring minimal information loss for internal data. Edge preserved kriging interpolation has been applied to obtain accurate shape and size preservation. The reconstruction of the sub parts were carried out in parallel in two phases. The time taken for reconstruction is approximately 56 seconds and that for virtual slicing in 2D is a fraction of a second. Viewing a cross sectional 3D internal tissue structure  takes approximately 25 seconds, depending on the number of slices required to get the desired cross section.  The proposed approach has been shown experimentally to work on MR images of human spine, brain and can be extended for any form of medical image with minor modification based on the technique of capture. In neuroscience, brain and spine are both of equal importance, thus we chose both these organs to validate our results.   Validation of this method on medical images of other types of organs, as a sequence of 2D slices can be taken up in future. This  method will help medical practitioners for pre-operative research as well as in virtual 3D volumetric modelling.

\section*{Acknowledgment}

The authors would like to thank Dr. Alok Pandit, Bangur Institute of Neurosciences, Kolkata, India for helping us with understanding of the real MR images and also validating our results.

\bibliographystyle{elsarticle-num}
\bibliography{bibliography}

\appendix

\section{ Kriging interpolation}\label{kriging}
Kriging \cite{25} is a spatial prediction technique that combines a regression of the dependent variable on auxiliary variables  with interpolation of the regression residuals. It is mathematically equivalent to the interpolation method variously called universal kriging and kriging with external drift, where auxiliary predictors are used directly to solve the kriging weights. 
In the case of kriging with external drift (KED), predictions at new locations are made by \cite{25}:

\begin{equation}
    \hat{z}_{KED}(s_0)=\sum_{i=1}^n w_i^{KED}(s_0) \cdot z(s_i)
\end{equation}

for

\begin{equation}
    \sum_{i=1}^n w_i^{KED}(s_0) \cdot q_k(s_i)= q_k(s_0)
\end{equation}

for $k=1, \dots, p$ or in matrix notation:
\begin{equation}
    \hat{z}_{KED}(s_0)=\delta_0^T \cdot z
\end{equation}

where $z$ is the target, $q_k$'s are the predictor variables $i.e.,$ values at a new location ($s_0$), $\delta_0$ is the vector of $KED$ weights ($w_i^{KED}$), $p$ is the number of predictors and $z$ is the vector of $n$ observations at primary locations. The $KED$ weights are solved using the extended matrices:

\begin{equation}
\resizebox{0.91\hsize}{!}{$
    \lambda_0^{KED}=\{w_1^{KED}(s_0),\dots, w_n^{KED}(s_0), \varphi_0(s_0),\dots, \varphi_p(s_0)\}^T= C^{KED-1} \cdot C_0^{KED}$}
\end{equation}

where $\lambda_0^{KED}$ is the vector of solved weights, $\varphi_p$ are the Lagrange multipliers, $C^{KED}$ is the extended covariance matrix of residuals and $c_0^{KED}$ is the extended vctor covariance at new location.

\section{Discrete Shearlet Transform}\label{shear}

A shearlet \cite{29} is generated by the dilation, shearing and translation of a function $\psi$ $\in$ $L^2 (\mathbb{R}^2)$, called the mother shearlet, in the following way \cite{24}

\begin{equation}
    \psi_{a,s,t}(x) = a^{-3/4}\psi(A^{-1}_a S^{-1}_s(x-t))
\end{equation}
where $t$ $\in$ $(\mathbb{R}^2)$
is a translation, $A_a$ is a scaling (or dilation)
matrix and $S_s$ a shearing matrix defined respectively by

\[
A_a = \begin{bmatrix} 
    a & 0 \\
    
    0 & \sqrt{a} 
    \end{bmatrix}
\qquad
S_s = \begin{bmatrix} 
    1 & -s \\
    
    0 & 1 
    \end{bmatrix}
\]

with $a \in (\mathbb{R}^+)$
 and  $s \in (\mathbb{R})$. The anisotropic dilation $A_a$ controls the scale of the shearlets, by applying a different dilation factor along the two axes. The shearing matrix $S_s$, not expansive, determines the orientation of the shearlets. The normalization factor $a^{-3/4}$ ensures that $\vert\psi_{a,s,t}\vert = \vert\psi\vert$, where $\vert\psi\vert$ is the norm in $L^2 (\mathbb{R}^2)$. In the classical setting the mother shearlet $\psi$ is assumed to
factorize in the Fourier domain as
\begin{equation}
\hat{\psi}(\omega_1, \omega_2) = \hat{\psi_1}(\omega_1)\hat{\psi_2}(\frac{
\omega_2}{\omega_1})
\end{equation}

where $\hat{\psi}$ is the Fourier transform of $\psi$, $\psi_1$ is a one dimensional wavelet and $\hat{\psi_2}$ is any non-zero square-integrable function. There are several examples of functions $\psi_1$, $\psi_2$ satisfying these
properties. The shearlet definition in the frequency domain:
\begin{equation}
\hat{\psi}_{a,s,t}(\omega_1, \omega_2) = a^{3/4}\hat{\psi_1}(a\omega_1)\hat{\psi}_2
(\frac{\omega_2 - s \omega_1}{\sqrt{a}\omega_1})
e^{-2\pi i(\omega_1,\omega_2)t}.
\end{equation}
The shearlet transform $SH(f)$ of a signal  $f$ $\in$ $L^2 (\mathbb{R}^2)$  is defined by

\begin{equation}
    SH(f)(a, s, t) = <f, \psi_{a,s,t}>
\end{equation}

where $<f, \psi_{a,s,t}>$ is the scalar product in $L^2 (\mathbb{R}^2)$. As a consequence of the Plancherel formula : 

\begin{equation}
\begin{split}
SH(f)(a, s, t) = a^{3/4}\int_{\hat{R}^2}
\hat{f}(\omega_1, \omega_2)\hat{\psi}_1(a\omega_1)\\\times \hat{\psi}_2
(\frac{\omega_2 - s \omega_1}{\sqrt{a}\omega_1}) \times e^{-2\pi i(\omega_1,\omega_2)}
d\omega_1 d\omega_2.
\end{split}
\end{equation}

\section{Marching Cube}\label{mc}
The Marching Cubes method \cite{15} is a simple iterative algorithm for creating a mesh of triangles to represent the surfaces of a given 3D object specified as a 3D array of pixels. The algorithm works by \emph{marching} over the entire image of the 3D object, which has been equally sub-divided into cubes. Each cube is called a voxel. The algorithm then determines whether the 3D image intersects a cube, and assigns boolean values to the corners of the cube accordingly. Intuitively, suppose the values at all the corners of the cube (i.e., the voxel) are 1. Then the cube is said to lie entirely inside the surface. Similarly, if all the corners of the cube have a value 0, then the cube is said to lie entirely outside the surface. In both cases, there would be no triangular surface passing through the cube. The main aim of the algorithm is to determine triangles (its intersection points, normals) in the cases where some of the corners of a cube are 1 and the others are 0. As there are 8 corners in a cube (voxel), there  are 256 cube configurations, which are stored in a look-up table. The final mesh is obtained through iterative linear interpolation. We have used the Marching cube algorithm \cite{20} for surface rendering part of the 3D reconstruction from the 2D slices of MRI.

\section{Similarity Metrics for 2D Images}\label{sm}
The definitions of three most popular similarity metrics for images that we have used for validating our results, are presented next. We have sliced out 2D images from the virtually reconstructed 3D image and have matched these with the ground-truth 2D images that we started with.

\subsection{ Root Mean Square Error}
The Root Mean Square Error ($RMSE$) \cite{16} is a frequently used measure of the differences between values predicted by an estimator and the values observed. It is the square root of the average of the square of the errors. $RMSE$ of an image $f_{1}(m,n)$ with respect to an image $f_{2}(m,n)$ is defined as the square root of the mean square error ($MSE$)\cite{19} :

\begin{equation*}
MSE=\dfrac{1}{MN}\sum_{n=1}^{N}\sum_{m=1}^{M}{[f_{2}(n,m)-f_{1}(n,m)]}^2
\end{equation*}

where $M\times N$ is the size of the image matrix. Thus, the 
$RMSE=\sqrt{MSE}$.
A value of $RMSE$ close to 0 implies that the probability of the two images being identical is higher.

\subsection{Mutual Information}
 Mutual information ($MI$) \cite{16} is a quantitative measure of information about one random variable $(Y)$ with respect to another random variable $(X)$. However, information is a reduction in the uncertainty of a variable. Hence, the higher is the mutual information between $X$ and $Y$, the lower is the uncertainty of $X$ given $Y$, or vice versa. 
 Let $G$ and $R$ be the ground-truth and the reconstructed images respectively. The mutual information $MI_{GR}$ between them is defined as:
\begin{equation}
MI_{GR}=\sum_{g,r}p_{G,R}(g,r)\log \frac{p_{GR}(g,r)}{p_G(g)p_R(r)}
\end{equation} 
where $p_{G,R}$ is the joint probability mass function of $G$ and $R$, $p_G$ and $p_R$ are the marginal probability mass function of $G$ and $R$, and $g$, $r$ represent the pixel value of image $G$ and image $ R$ respectively. 

$MI$ has been used for checking the accuracy of the generated missing data, compared to the available ground truth data. The greater is the $MI$ between a generated slice and the original  slice, the better is the accuracy of our reconstruction algorithm.

\subsection{Structural Similarity Index Method}
Structural Similarity Index Method ($SSIM$) \cite{19} is quantifies the similarity between two images. The measurement or prediction of image quality is based on an initial uncompressed or distortion-free image as reference. The $SSIM$ index between two images $X$ and $Y$  is obtained as:

\begin{equation*}
SSIM(X,Y)=\dfrac{(2\mu_X\mu_Y+c_1)(2\sigma_{XY}+c_2)}{(\mu_X^2+\mu_Y^2+c_1)(\sigma_X^2+\sigma_Y^2+c_2)}
\end{equation*}

where $\mu_X$ and $\mu_Y$ are the mean, $\sigma_X^2$ and $\sigma_Y^2$ are the standard deviation  of $X$ and $Y$ respectively. $\sigma_{XY}$ is the covariance of $X$ and $Y$, and the constants $c_1$ and $c_2$  stabilize the ratio with a weak denominator. $SSIM = 1$ implies that the two images can be considered to be identical.

\end{document}